\documentstyle[prl,aps,twocolumn]{revtex} 
\begin{document}
\draft
\reversemarginpar
\title{Vortex dynamics in layered superconductors with correlated defects:
influence of interlayer coupling}
\author{A. Engel\cite{byline},$^{1}$ H. J. Trodahl,$^{1}$ J. C. Abele,$^{2}$ D. Smith,$^{2}$ and S. M. Robinson$^{2}$}
\address{$^{1}$School of Chemical and Physical Sciences, Victoria University of Wellington, Wellington, New Zealand\\
$^{2}$Physics Department, Lewis \& Clark College, Portland, Oregon 97219}
\maketitle
\begin{abstract}
We report a detailed study of the vortex dynamics and vortex phase diagrams of
two amorphous Ta$_{0.3}$Ge$_{0.7}$/Ge multilayered films with intrinsic
coplanar defects, but different interlayer coupling. A pinned Bose-glass phase
in the more weakly coupled sample exists only below a cross-over field $H^{*}$
in striking contrast to the strongly coupled film. Above $H^{*}$ the flux
lines are thought to break up into pancake vortices and the cross-over field is
significantly increased when the field is aligned along the extended defects.
The two films show different vortex creep excitations in the Bose-glass phase.
\end{abstract}

The magnetic phase diagrams of type-II superconductors show a rich variety of
magnetic vortex phases and phase-transitions in the mixed state
\cite{Blatter94}. The second order melting transition from a highly
disordered, but pinned vortex glass to an unpinned vortex liquid is one of
considerable interest since the solid glass phase shows zero resistance in the
limit of small applied currents. The field and temperature dependencies of the
glass melting line are influenced by thermal fluctuations, quenched disorder,
electronic anisotropy and dimensionality; furthermore the defects'
dimensionality, point-like (0D) or extended (1D, 2D), influences the nature of
the transition and glass phase \cite{Fisher89,Fisher91,Nelson93}. Many of
these points have been addressed and compared in strongly-coupled
YBa$_{2}$Cu$_{3}$O$_{7-\delta}$ (YBCO) and the very weakly-coupled
Bi$_{2}$Sr$_{2}$CaCu$_{2}$O$_{8+\delta}$ (BSCCO), but a complete understanding
of the phase diagrams and the role of the different parameters will benefit
from studies in materials which permit continuous control of the
superconducting and layering parameters.

In this letter we report results on two amorphous multilayered films of
superconducting Ta$_{0.3}$Ge$_{0.7}$ alloy and insulating Ge with different
interlayer coupling which results in quite different phase diagrams. These
films are ideal systems to study the influence of dimensionality and
anisotropy on the vortex phase diagram. They belong to the strongly type-II
superconductors ($\lambda/\xi \approx 100$), layer thickness can be varied
continuously over a wide range and oblique vapor deposition \cite{Smith60}
allows the introduction of intrinsic extended defects spanning the whole film
thickness \cite{Abele99,Engel01}. Additionally, the geometry of the film
microstructure allows the comparison of two otherwise identical magnetic field
configurations: one with the field parallel to the defects when the flux lines
can take full advantage of their extended nature and another, symmetrically
across the film normal, at large angles to the defects.

Details of the film preparation have been described earlier
\cite{Ruck98a,Engel01}. They were deposited under UHV conditions and with the
substrate tilted by $40^{\circ}$ from the usual normal incident direction.
This results in completely amorphous multilayers with a columnar
microstructure, leaning columns of normal density are separated by a network of
reduced density. The defect structure is comparable to grain boundaries or
twin planes in twinned YBCO. TEM photographs of the cross-section of the films
reveal the layered nature as well as the direction of the columns
\cite{Engel01}. In Table \ref{tab1} the layer thickness and columnar angle as
derived from such photographs are given. Sample C40 is the strongly coupled
film with an insulator (i) thickness of $3.5$ nm, compared to $7.5$ nm for
sample P40; the superconducting (sc) layer thicknesses in the two films are
approximately the same at $13.5$ nm and $13.0$ nm, respectively. The leaning
angle of the columns measured from the film normal is $19.5^{\circ}$ and
$25.5^{\circ}$ and the number of sc/i bi-layers is $12$ and $16$ for C40 and
P40, respectively.

DC, 4-point conductivity measurements were chosen to examine the
superconducting behavior of the films. All measurements were made with the
sample immersed in liquid helium providing thermal stability even at high
applied current densities. Temperatures were measured using a carbon-glass
resistor in close proximity and good thermal contact with the sample mounting
block. The achieved temperature stability was better than $1$ mK below the
lambda point and $\approx 3$ mK above, at an absolute accuracy of the order of
$10$ mK. External currents were applied in the film plane and perpendicular to
the columnar film structure and applied magnetic fields, as illustrated in the
insert of Fig.\ \ref{fig1}. The magnetic field could be rotated in the plane
perpendicular to the current.

The variation of the resistance with magnetic field orientation for fixed $H$
and $T$ is shown for both films in Fig.\ \ref{fig1}. Clearly visible is a
local minimum at about $20^{\circ}$ -- $30^{\circ}$ for fields parallel to the
columnar orientation ({\it C} direction). There is a systematic deviation of
columnar angles determined from the resistance minimum and those from the TEM
photographs (see Table \ref{tab1}); the latter are always a few degrees
smaller. The TEM photographs were not available at the time of the conductance
measurement so that resistance were used to define columnar direction. Here we
focus on that and the anti-columnar ({\it AC}) direction at the same angle to
but on the opposite side of the film normal, where the extended nature of the
defects is ineffective.

Zero-field $R$-$T$ measurements show a sharp superconducting transition with
no steps or other irregularities pointing to the presence of inhomogenities.
In-field data give a linear $H_{c2}(T)$ for the {\it C} and {\it AC}
directions with extrapolated values at zero temperature given in table
\ref{tab1} together with other key parameters. Information about the
glass-liquid transition was derived from sets of $IV$-characteristics taken at
constant magnetic field and temperatures in the vicinity of the melting
temperature. As pointed out by \cite{Fisher89,Fisher91} and \cite{Nelson93} in
the case of point and correlated disorder, respectively, such $IV$-curves
should show universal scaling behavior. The relation between current density
$J$ and electric field $E$ can be summarized as follows:
\begin{equation}
\left(E/J\right) \left|t\right|^{\nu\left(D-2-z\right)} \propto F_{\pm}\left(J
\left|t\right|^{-\nu\left(D-1\right)}\right),
\end{equation}%
with $\nu$ the static and $z$ the dynamic critical exponent,
$t=1-\left(T/T_{m}\right)$ the reduced temperature and $T_{m}$ the melting
temperature. The parameter D is 3 or 4, respectively, within the 3D-vortex
(point disorder) or Bose-glass (correlated disorder) theories. Above $T_{m}$
and for small currents $F_{+}(x \rightarrow 0)=\textrm{const.}$, indicating
ohmic resistance. Below $T_{m}$ in the glass phase the resistance vanishes
exponentially with $F_{-}(x \rightarrow 0)=\exp(x^{-\mu})$.

Strongly coupled sample: sample C40 shows the typical change for a glass
transition from ohmic curves with a positive curvature at high currents and
temperature to downwards curved $IV$-characteristics below the melting
temperature. Scaling analysis has been applied successfully using Bose-glass
scaling laws for $IV$-sets taken with the field applied in the {\it C} and
3D-vortex glass laws for the {\it AC} direction. The resulting parameters are
given in Table \ref{tab1}. The phase diagram Fig.\ \ref{fig2} shows a
significantly increased glassy phase region for fields along {\it C} due to
increased pinning by extended defects.

Weakly coupled sample: here, the situation is completely different. Of all the
$IV$-sets taken with the magnetic field in the {\it C} direction, the
characteristic glass-melting behavior was observed and scaling was successful
only at the lowest field ($0.1$ T). At higher fields, the $IV$-curves show no
sign of negative curvature, in fact they have positive curvature even at the
lowest temperatures, as illustrated in Fig.\ \ref{fig3}. Vortex glass theory
for two-dimensional superconductors lacks the transition into a
low-temperature zero-resistance state \cite{Fisher91}. Thus, our
interpretation is that there is a phase transition from a vortex liquid into a
Bose-glass at low magnetic fields only, and that there is another transition
into a 2D or quasi-2D liquid at higher fields which then does not show any
glassy behavior. This view is further supported by the $IV$-curves taken at
the second lowest field ($0.3$ T, Fig.\ \ref{fig3}). At high temperatures we
see the usual ohmic curves with an upturn at higher currents, but at lower
temperatures all the curves show approximately power-law behavior. Since the
power-law curve determines the melting temperature, this field $H^{*}=0.3$ T
may correspond to the cross-over field from 3D to quasi-2D behavior
(horizontal line in Fig.\ \ref{fig4}). Thus we have observed field-driven 3D
to 2D transitions which have previously been reported only in cuprates
\cite{Schilling93}.

This behavior contrasts with that in {\it AC} aligned fields, for which the
defects are uncorrelated. In this geometry the stability of a 3D phase is
reduced further, so that no evidence of glassy behavior is seen even at the
lowest measured field. However, at this lowest field the vortex system is very
close to the proposed dimensional transition, as evidenced by the observation
that all $IV$-curves below $1.95$ K show nearly ideal power-law dependence. The
implied phase boundaries are then given as the dotted lines in Fig.\
\ref{fig4}. Thus for {\it C} oriented fields the extended defects preserve the
integrity of the flux lines and promote the 3D phase, confirming a suggestion
based on recent measurements on the cuprate superconductor BSCCO.
\cite{Morozov99,Sato97}

Recently, Ammor et al.\ \cite{Ammor00} published detailed data on BSCOO with
columnar defects and it is instructive to compare the resulting phase diagrams
to ours. In that study the Bose-glass phase was found to be limited to fields
smaller than the matching field $B_{\Phi}$ above which the vortices outnumber
the columnar defects. Thus the Bose-glass phase disappears at fields large
enough that some vortices remain unpinned. Though it is not straight forward
to define a matching field in the case of coplanar defects, we see a maximum
increase in activation energy in the thermally activated flux flow region
between $0.5$ and $1.0$ T. These fields are a factor $2$ to $3$ higher than
the observed cross-over field in the $IV$-characteristics. Hence, the cause
for the difference in $IV$-curves above and below $H^{*}=0.3$ T cannot be
attributed to the occurrence of unpinned flux lines.

Analysis of the Bose-glass phase reveals further differences between the
strongly and weakly coupled sample. There is an obvious difference in the
glass parameters $z$ and especially $\nu$ resulting from different mechanisms
for flux creep in the glass phase of the two samples. For low current
densities flux creep should be dominated by variable-range-hopping processes
which give rise to the following relation between electric field $E$ and
current density $J$ \cite{Nelson93}:
\begin{equation}\label{VRH}
E/J \propto \exp\left[-\alpha\left(J_{0}/J\right)^{\mu}\right],
\end{equation}%
with $J_{0}$ a characteristic current scale, $\alpha$ a temperature dependent
factor including a characteristic energy for vortex kinks and $\mu$ the glass
exponent; typical values for dominant excitations are $\mu=1$ for vortex-loops
and $\mu=1/3$ for double-superkinks, respectively. Replotting the $IV$-curves
in the Bose-glass phase by $\rm{d}[\ln(V/I)]/\rm{d}I$ versus $I$ should result
in a straight line with gradient $-(\mu+1)$. In Fig.\ \ref{fig5} examples for
both films are given. Whereas the low current part for the weakly-coupled
sample P40 can be fitted well with $\mu=1/3$, data for C40 can only be fitted
with $\mu=1$. Clearly the energy scale for vortex kinks is much higher in the
strongly coupled film C40 and the excitation of vortex loops is needed for
flux creep. In contrast the excitation of kinks is easier in the weakly
coupled sample, where segments of the flux line, i.e.\ individual pancake
vortices, can jump to the next favorable pinning site through the generation
of a pair of superkinks,  giving rise to the observed exponent $\mu = 1/3$.

In summary, we have observed striking contrasts in the vortex phase diagrams of
two extremely type-II superconducting multilayered films of
Ta$_{0.3}$Ge$_{0.7}$/Ge with very similar coplanar defects but different
interlayer coupling. The more strongly coupled film shows a conventional phase
diagram, with an extended region of vortex solid phases, but in the weakly
coupled sample a strongly pinned Bose-glass phase exists only for weak
magnetic fields aligned with the defect structure. Above a relatively small
cross-over field $H^{*}$ the glass phase melts into a quasi-2D liquid of
pancake vortices. Correlated defects increase the solid-phase region , but
$H^{*}$ is severely reduced, possibly to zero, in magnetic field orientations
for which the extended nature of the defects is ineffective. The Bose-glass
phases in the two films also differ in the dominant mode by which their low
current flux creep proceeds. Whereas flux creep in the weakly coupled film can
be described well by the generation of superkinks, the creep process for the
more rigid flux lines in the strongly coupled sample is dominated by the
excitation of vortex loops.

The authors wish to thank Prof.\ P.\ Munroe from the Electron Microscope Unit
at the University of New South Wales, Australia, for the TEM photographs and
acknowledge support by the New Zealand Marsden Fund and the Research
Corporation Cottrell College Science Award (CC4444).


\clearpage
\begin{table}
\caption{Various parameters for both films. $T _{c}(0)$  determined from
zero-field resistance data and $H _{c2}(0)$ from scaling in-field data
according to fluctuation conductivity theory 
. The penetration
depth and coherence length are determined respectively, by $\lambda (0)
\approx 1.05 \times 10^{-3}(\rho _{0}/T _{c}(0))^{1/2}$ and $\xi (T) = (\Phi
_{0}/[2\pi H _{c2}(T)])^{1/2}$, $\rho _{0}$ normal state resistivity at $T=0$
and $\Phi _{0}$ the flux quantum. Individual layer thicknesses $d$(sc) and
$d$(i) and columnar orientation $\beta$(TEM) were determined from TEM images.
$\beta$(TEM) is compared to $\beta$(rot) derived from resistance measurements
(see Fig.\ \ref{fig1}). The last two columns give the glass parameters for
{\it C} and {\it AC} directions using Bose-glass and 3D-vortex glass scaling,
respectively. All lengths given in nm.} \label{tab1}
\begin{tabular}{lcccccccc}
 & \multicolumn{1}{c}{$T_{c}(0)$ [K]} & \multicolumn{1}{c}{$H_{c2}(0)$ [T]} & \multicolumn{1}{c}{$\lambda$} & \multicolumn{1}{c}{$\xi$} & \multicolumn{1}{c}{$d$(sc/i)} & \multicolumn{1}{c}{$\beta$(TEM/rot)} & \multicolumn{1}{c}{$z$ (C/AC)} & \multicolumn{1}{c}{$\nu$ (C/AC)}\\\tableline
C40 & $2.395$ & $7.4$ & $1300$ & $7$ & $13.5/3.5$ & $19.5^{\circ}/23.3^{\circ}$ & $8.91/5.42$ & $0.69/1.24$ \\
P40 & $2.272$ & $7.3$ & $1500$ & $7$ & $13.0/7.5$ & $25.5^{\circ}/27.4^{\circ}$ & $7.44/-$ & $1.27/-$ \\
\end{tabular}
\end{table}

\clearpage

\begin{figure}[h!p]
\caption{Resistance versus magnetic field orientation for C40 (taken at
$2.00$ K and $0.4$ T) and P40 ($1.90$ K/$0.3$ T). The films' anisotropy is
clearly visible with the second minimum for fields directed parallel to the
defect structure. The measurement geometry is shown schematically in the upper
right corner. The angle is measured from the film normal with positive values
towards the {\it C} direction.} \label{fig1}
\end{figure}

\begin{figure}[h!p]
\caption{Reduced $H$-$T$ phase diagram for C40 determined from scaling
analysis. The second order phase transition from a pinned glass phase to the
vortex liquid phase shifted to higher temperatures and fields for fields in
the {\it C} (squares) compared to the {\it AC} direction (triangles).}
\label{fig2}
\end{figure}

\begin{figure}[h!p]
\caption{Logarithmic gradient $\rm{d}(\ln V)/\rm{d}(\ln I)$ versus $I$ of
$IV$-curves taken for P40 at different applied magnetic fields in the {\it C}
direction and varying temperatures ($\Delta T=50$ mK). At $0.1$ T the sample
shows the typical glass-to-liquid transition from upward to downward
curvature. At $0.6$ T even the curves at the lowest temperatures (uppermost
lines) show increasing gradients. At the intermediate field of $0.3$ T most
curves exhibit power-law behaviour which is identified with a decoupling
transition from vortex lines to pancake vortices.} \label{fig3}
\end{figure}

\begin{figure}[h!p]
\caption{Reduced $H$-$T$ phase diagram for P40. The Bose-glass exists only
below the solid horizontal line (C) and melts into a 3D-liquid (which may be a
mixture of 3D- and 2D-liquids
). The dashed line labeled AC
indicates the glass phase boundaries for magnetic fields oriented in the {\it
AC} direction. $H_{max}$ is the highest applied field.} \label{fig4}
\end{figure}

\begin{figure}[h!p]
\caption{Log-log plot of $\rm{d}[\ln(V/I)]/\rm{d}(I)$ versus $I$ for $T<T_{m}$
and field in the {\it C} direction. Open symbols are results for P40 taken at
$0.1$ T and $1.60$ K (triangles) and $1.90$ K (circles), respectively. The
solid lines are fits according to Eq.\ \ref{VRH} with $\mu=1/3$. Filled
symbols are results for C40 taken at $0.3$ T and $1.55$ K (triangles) and
$1.80$ K (circles) (for easier separation from P40 data are plotted versus
$10\times I$). For C40 satisfactory fitting could only be achieved with $\mu=1$
(solid lines; compare to the dotted lines with $\mu=1/3$).} \label{fig5}
\end{figure}


\begin{references}
\bibitem[*]{byline} corresponding author. School of Chem.\ and Phys.\ Sciences,
Victoria University of Wellington, P.O. Box 600, Wellington, New Zealand.
Email address: andreas.engel@vuw.ac.nz

\bibitem{Blatter94}
G. Blatter {\it et al.}, Rev.\
  Mod.\ Phys. {\bf 66},  1125  (1994).

\bibitem{Fisher89}
M.~P.~A. Fisher, Phys.\ Rev.\ Lett. {\bf 62},  1415  (1989).

\bibitem{Fisher91}
D.~S. Fisher, M.~P.~A. Fisher, and D.~A. Huse, Phys.\ Rev.\ B {\bf 43},  130
  (1991).

\bibitem{Nelson93}
D.~R. Nelson and V.~M. Vinokur, Phys.\ Rev.\ B {\bf 48},  13060  (1993).

\bibitem{Smith60}
D.~O. Smith, M.~S. Cohen, and G.~P. Weiss, J.\ Appl.\ Phys. {\bf 31}, 1755
  (1960);  A.~G. Dirks and H.~J. Leamy, Thin\ Solid\ Films {\bf 47},  219
  (1977);  H. van Kranenburg and C. Lodder, Mat. Sci. and Engineering {\bf R11}, 295  (1994).

\bibitem{Abele99}
J.~C. Abele {\it et al.}, Phys.\ Rev.\ B {\bf 60},  12448  (1999).

\bibitem{Engel01}
A. Engel, H.~J. Trodahl, J.~C. Abele, and S.~M. Robinson, Phys.\ Rev.\ B {\bf
63}, 184502 (2001).

\bibitem{Ruck98a}
B. Ruck, Ph.D. thesis, {V}ictoria {U}niversity of {W}ellington, {W}ellington,
  {N}ew {Z}ealand, 1998.

\bibitem{Ullah90}
S.~Ullah and A.~T.~Dorsey,
  Phys.\ Rev.\ Lett. {\bf 65}, 2066, (1990); Phys.\ Rev.\ B {\bf 44}, 262, (1991).

\bibitem{Schilling93}
A. Schilling, R. Jin, J.~D. Guo, and H.~R. Ott, Phys.\ Rev.\ Lett. {\bf 71},
  1899  (1993); H. Obara {\it et al.}, {\it ibid.} {\bf 74},  3041  (1995);
  F. Zuo, S. Khizroev, G.~C. Alexandrakis, and V.~N. Kopylov, Phys.\ Rev.\ B {\bf 52},  R755  (1995).

\bibitem{Morozov99}
N. Morozov {\it et al.}, Phys.\ Rev.\ Lett. {\bf 82},  1008  (1999); N.
Morozov, M.~P. Maley, L.~N. Bulaevskii, and J. Sarrao, Phys.\ Rev.\ B {\bf
57},  R8146  (1998); M. Kosugi {\it et al.}, Phys.\ Rev.\ Lett. {\bf 79},  3763
(1997).

\bibitem{Sato97}
M. Sato {\it et al.}, Phys.\ Rev.\
  Lett. {\bf 79},  3759  (1997).

\bibitem{Ammor00}
L. Ammor {\it et al.}, J.\ Phys.:\ Condens.\ Matter {\bf 12},  4217  (2000).

\end{references}
\end{document}